\documentclass{webofc}

\usepackage[varg]{txfonts}   
\usepackage{hyperref}
\usepackage{url}
\hypersetup{colorlinks=true,citecolor=blue,urlcolor=blue,linkcolor=blue}
\newcommand{\fs}{60\,\mathrm{MSPS}}

\newcommand{\diagramwidth}{0.96\linewidth}
\newcommand{\wideplotwidth}{0.86\linewidth}
\newcommand{\plotwidth}{0.80\linewidth}

\begin{document}

\title{Mitigating Deadtime in Distributed Optical Arrays\\ 
Using A Liveness-Aware Trigger Approach\\ 
for High-Energy Neutrino Detection}

\subtitle{%
\small\normalfont
Oral presentation in Track 2: Online \& Real-Time Computing\\
28th Conference on Computing in High Energy and Nuclear Physics (CHEP 2026)\\[2pt]
\textit{Prepared for submission to EPJ Web of Conferences}
}

\author{\firstname{Thammarat} \lastname{Yawisit}\inst{1,2}\fnsep\thanks{Corresponding author: \email{65010454@kmitl.ac.th}; \url{tyawisit@icecube.wisc.edu}}
        \and
        \firstname{Pittaya} \lastname{Pannil}\inst{1}}

\institute{Department of Instrumentation and Control Engineering, School of Engineering,\\ King Mongkut's Institute of Technology Ladkrabang, Bangkok, Thailand 10520
\and
Princess Srisavangavadhana Faculty of Medicine,\\ Chulabhorn Royal Academy, Bangkok, Thailand 10210}

\abstract{Large-scale neutrino observatories operate under unavoidable detector deadtime arising from photomultiplier saturation, digitizer limits, and front-end readout constraints. Conventional coincidence-based trigger logic implicitly assumes continuous sensor availability and therefore suffers systematic efficiency loss when channels become temporarily non-live. This work presents the design of a liveness-aware trigger architecture targeting low-latency FPGA deployment in distributed optical arrays. We introduce a recursive Infinite Impulse Response (IIR) update law implemented as a fully synthesizable pipeline that constructs a continuity-preserving effective observable at each sensor node. Rather than collapsing during non-liveness intervals, the observable decays smoothly while retaining phase and amplitude information relevant for network-level coherence estimation. By explicitly separating continuous measurement construction from discrete trigger decision logic, the proposed architecture enables graceful degradation under partial channel non-liveness. Simulation results demonstrate sustained event recovery efficiency in regimes of elevated deadtime probability, where conventional coincidence logic degrades substantially.}

\maketitle

\clearpage
\section{Introduction}
\label{sec:introduction}

High-energy neutrino observatories rely on large-scale distributed optical sensor
arrays to detect faint Cherenkov radiation produced by neutrino interactions in
transparent media such as Antarctic ice
\cite{IceCubeScience,IceCubeDetector}.
Similar detection principles are employed by other cubic-kilometer--scale
experiments in different environments, including seawater and lake ice, such as
KM3NeT and Baikal-GVD, underscoring the generality of distributed optical arrays as
a detection paradigm for neutrino astronomy
\cite{KM3NeT,BaikalGVD}.
Because astrophysical neutrino events are intrinsically rare, data acquisition
(DAQ) and trigger systems must operate continuously in the presence of
overwhelming backgrounds arising from atmospheric muons, thermal and radioactive
optical noise, and electronics artifacts
\cite{IceCubeTrigger}.
The primary role of the trigger system is therefore to preserve sensitivity to
correlated neutrino-induced signals while controlling bandwidth, latency, and
false trigger rates, a challenge shared across modern high-energy and
multimessenger observatories
\cite{MultiMessenger,DAQReview}.

\begin{figure}[!h]
\centering
\begin{minipage}[b]{0.32\textwidth}
  \centering
  \includegraphics[width=\linewidth]{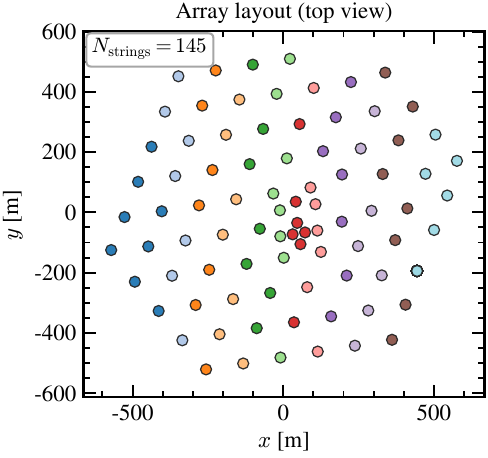}\\[-0.3ex]
  \small (a) Top view ($xy$-plane)
\end{minipage}\hfill
\begin{minipage}[b]{0.32\textwidth}
  \centering
  \includegraphics[width=\linewidth]{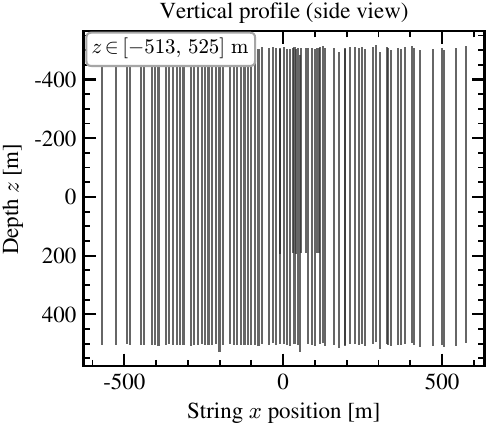}\\[-0.3ex]
  \small (b) Side view ($xz$-plane)
\end{minipage}\hfill
\begin{minipage}[b]{0.32\textwidth}
  \centering
  \includegraphics[width=\linewidth]{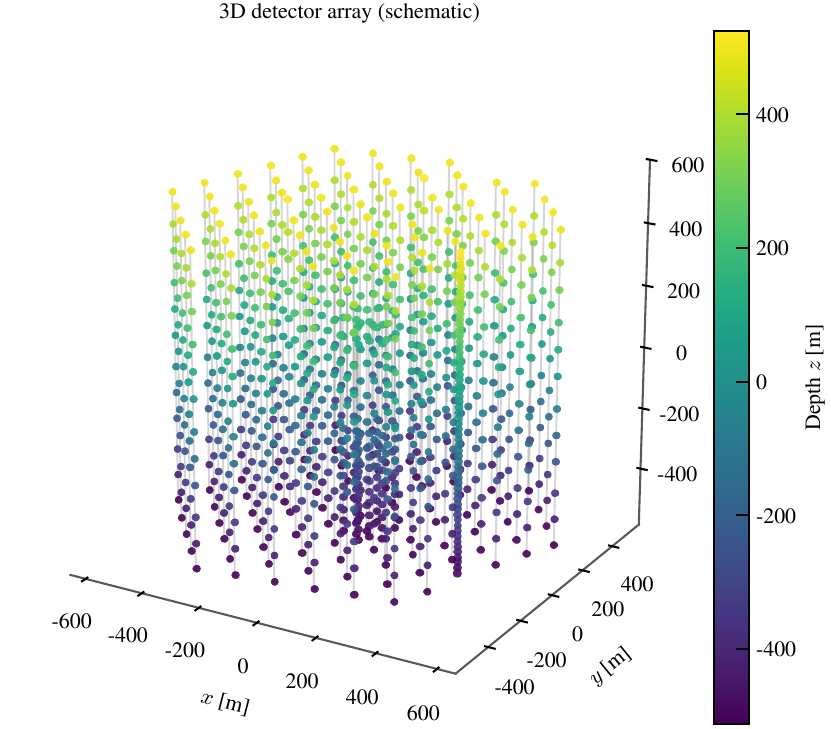}\\[-0.3ex]
  \small (c) 3D isometric view
\end{minipage}
\caption{Three complementary views of a representative distributed optical detector geometry: (a) the horizontal layout in the $xy$-plane, (b) the vertical deployment profile in the $xz$-plane, and (c) a 3D isometric rendering illustrating the overall footprint and depth extent.}
\label{fig:array_geometry_multiview}
\end{figure}

Conventional trigger architectures in distributed optical arrays are predominantly
based on binary coincidence logic, including local coincidence between neighboring
sensors and global multiplicity conditions across the detector volume.
Under nominal operating conditions, coincidence-based triggers provide a robust
and computationally efficient mechanism for event selection
\cite{IceCubeTrigger}.
However, such architectures implicitly assume that all sensors are continuously
live and capable of reporting reliable hit information.

In practice, detector deadtime is not an exceptional corner case but an intrinsic
property of realistic optical sensors and their front-end electronics.
Deadtime arises from multiple physical and electronic constraints, including
photomultiplier tube (PMT) saturation under large light deposits,
analog-to-digital converter (ADC) clipping, finite waveform readout buffers,
and local reset or veto logic.
From a counting and detection-theory perspective, these effects correspond to
well-established non-paralyzable or mixed deadtime processes that lead to
systematic information loss at high rates
\cite{DeadtimeKnoll,NonParalyzable}.
During such intervals, a sensor becomes temporarily non-live and cannot contribute
valid information to the trigger decision.
In coincidence-based architectures, non-live sensors are effectively interpreted
as the absence of a signal, thereby breaking otherwise valid coincidence chains
and resulting in a systematic reduction in detection efficiency.

These limitations become increasingly relevant for next-generation large-scale
detector systems that incorporate heterogeneous sensor responses, wide dynamic
ranges, and on-node digital processing.
While deadtime effects can often be mitigated during offline reconstruction, they
are not naturally accommodated by binary coincidence logic implemented at the
firmware trigger level.
This mismatch motivates the development of trigger architectures that explicitly
account for partial sensor non-liveness while preserving sensitivity to
temporally correlated signals.

In this work, we propose a liveness-aware trigger architecture in which the
distributed sensor network is treated as a coherent measurement system whose
internal state evolves continuously, even when a subset of nodes is temporarily
non-live.
Rather than encoding trigger conditions purely as discrete coincidence windows,
the proposed approach constructs a continuity-preserving effective observable at
each sensor node using a recursive Infinite Impulse Response (IIR) update law.
This formulation enables smooth temporal evolution of the observable during
non-liveness intervals while retaining phase and amplitude information relevant
for subsequent network-level coherence estimation.

The proposed trigger formulation builds upon the Synchromodulametry framework
\cite{Yawisit:2025ipb}, which treats coherence as a hardware-native state variable
derived from causal filtering, metric-aware timing alignment, and correlation-based
measures.
Performance validation is conducted using a hybrid framework that combines
representative event topologies derived from IceCube Open Data
\cite{IceCubeOpenData}
with synthetic signal injection and controlled deadtime stress tests.
This strategy enables systematic evaluation of trigger performance under
increasing deadtime while preserving physically meaningful event structure.

\subsection{Contributions and Paper Organization}
This paper makes three main contributions:
\begin{itemize}
  \item the formulation of a liveness-aware coherence observable suitable for
  real-time digital signal processing and FPGA deployment;
  \item a validation methodology that couples real event topologies with
  controlled deadtime injection and hardware-realistic signal modeling; and
  \item quantitative results demonstrating improved event recovery under
  increasing deadtime when compared with baseline coincidence-based triggering.
\end{itemize}

The remainder of this paper is organized as follows.
Section~\ref{sec:methods} describes the system architecture and analysis methods.
Section~\ref{sec:results} presents the trigger validation and performance results.
Section~\ref{sec:discussion-conclusion} discusses the implications of the proposed approach and concludes the
study.

\clearpage
\section{Materials and Methods}
\label{sec:methods}

\subsection{System Overview}

\begin{figure}[!h]
\centering
\includegraphics[width=\diagramwidth]{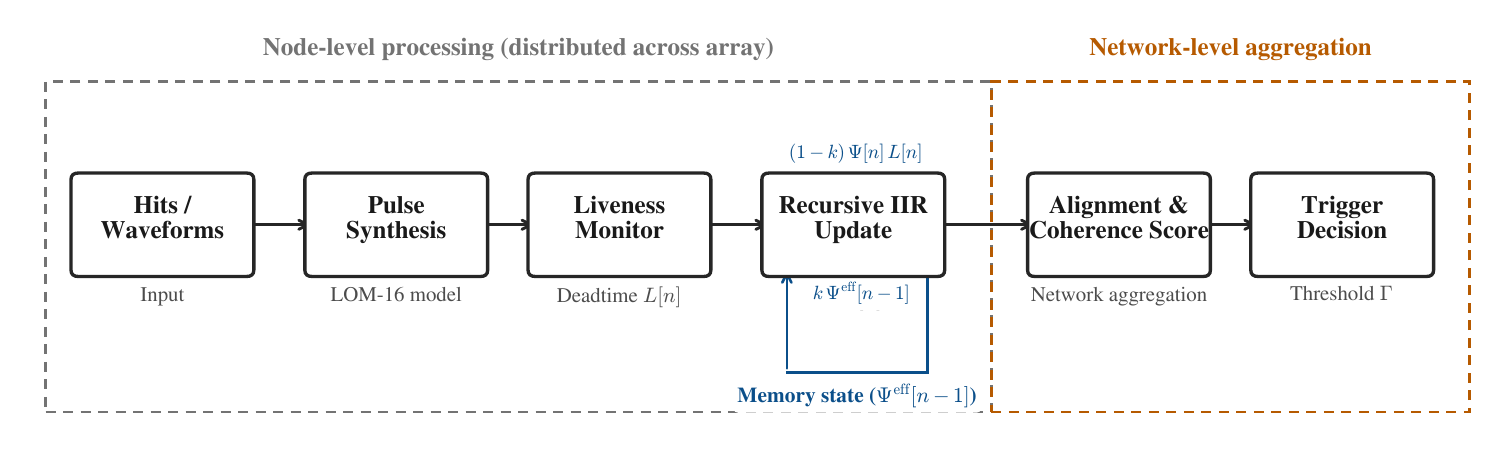}
\caption{Pipeline architecture of the proposed liveness-aware trigger system.}
\label{fig:block}
\end{figure}

The proposed trigger system is implemented as a modular, streaming pipeline
illustrated in Fig.~\ref{fig:block}, following standard real-time DSP design
principles for continuous, low-latency stream processing
\cite{StreamingDSP}.
The processing chain consists of
(i) event and hit ingestion, (ii) pulse synthesis,
(iii) liveness detection and masking,
(iv) recursive effective observable update,
(v) multi-node alignment and coherence scoring, and
(vi) trigger decision logic.
A key architectural choice is the explicit separation between
\emph{measurement}, the continuous construction of a physically meaningful
per-channel observable, and \emph{decision}, thresholding and window-based
logic used to generate triggers. This separation is a standard robustness
principle in real-time trigger design: it allows the measurement layer to
remain stable and information-preserving even when the decision layer is tuned
conservatively for bandwidth and false-rate constraints
\cite{DAQReview,FPGA_DSP}.

Each stage operates causally on streaming data and maintains only minimal local
state, avoiding global synchronization and enabling scaling to large channel
counts.
In particular, liveness awareness is introduced \emph{before} network-level
aggregation so that downstream coherence scoring can distinguish between
measurement unavailability and true silence.
As a result, the pipeline is designed to degrade gracefully under partial
non-liveness rather than collapsing via broken coincidence chains.

\subsection{Temporal and Computational Constraints}
The trigger system is designed under strict real-time constraints imposed by
high-rate data acquisition and limited on-detector computational resources.
In large-scale optical arrays, trigger latency directly impacts buffering
requirements, data throughput, and downstream processing efficiency
\cite{IceCubeTrigger,RealTimeTrigger}.

All operations in the proposed pipeline are causal and streaming, requiring
only local state and a fixed-size sliding window.
The recursive IIR update (introduced in Sec.~2.5) has constant-time complexity
$\mathcal{O}(1)$ per sample per channel and requires only one multiply-accumulate
plus lightweight liveness control, making it suitable for FPGA fabric or
embedded DSP cores \cite{FPGA_DSP}.
Likewise, the coherence score in Eq.~(\ref{eq:coherence_energy}) scales linearly
with the number of participating channels and avoids matrix inversion or
iterative solvers, requiring no global synchronization beyond the window
boundary.

These constraints ensure that the proposed trigger can operate continuously at
full detector rate without introducing additional deadtime or unpredictable
latency.
From a hardware perspective, the absence of global buffers, large state
machines, or iterative computations supports deterministic resource usage and
timing closure in low-latency firmware pipelines \cite{LowLatencyFPGA,FirmwareDAQ}.
All operations are compatible with fixed-point arithmetic and bounded dynamic
range, enabling implementation on low-power FPGA families such as Lattice
iCE40UP or similar embedded platforms \cite{FixedPointDSP}.

\subsection{Parametric Signal and Readout Model}
To ensure that trigger evaluation reflects realistic detector behavior while
remaining experiment-agnostic, the pulse synthesis stage employs a parametric
photomultiplier tube (PMT) response and digitizer model representative of modern
optical sensor front ends.
The model captures three essential features:
(i) the single-photoelectron (SPE) impulse response,
(ii) gain scaling and charge accumulation, and
(iii) saturation and clipping effects arising from finite digitizer dynamic
range.

Deadtime mechanisms are modeled explicitly as non-liveness windows associated
with ADC overflow, waveform buffering limits, or readout recovery intervals.
These abstractions reflect common operational constraints in distributed optical
modules and enable controlled stress testing at the trigger level.

A commonly adopted analytic approximation for the SPE pulse shape is the
difference-of-exponentials model \cite{OppenheimDSP,ProakisDSP}:
\begin{equation}
V(t) = G \cdot \frac{1}{\tau_d - \tau_r}
\left(e^{-t/\tau_d} - e^{-t/\tau_r}\right), \quad t \ge 0,
\label{eq:spe}
\end{equation}
where $G$ is the gain factor and $\tau_r$ and $\tau_d$ denote the rise and decay
time constants, respectively.

In addition to the analytic SPE shape, the model includes channel-dependent gain
and baseline-noise variations to emulate realistic non-uniformities across
multi-channel sensor nodes (e.g., manufacturing tolerances, high-voltage
differences, and temperature-dependent electronics behavior)
\cite{IceCubeDetector}.

Saturation effects are modeled by imposing a hard ceiling corresponding to the
ADC dynamic range, followed by a recovery interval during which the channel is
marked non-live.
This representation captures the dominant trigger-level impact of saturation
while avoiding unnecessary complexity in the simulation stage.

\begin{table}[!h]
\centering
\caption{Key parameters used in the parametric signal and readout model.}
\begin{tabular}{l c}
\hline
Parameter & Value / Description \\
\hline
Sampling rate & $\fs$ \\
ADC resolution & 12-bit (0--4095 counts) \\
Waveform readout window & $1.5~\mu$s \\
Baseline (pedestal) & $\sim$300 ADC counts (representative) \\
PMT gain $G$ used for PE conversion & representative value (order $10^6$--$10^7$ e$^-$/PE) \\
Charge per 1 PE ($Q_{1PE}$) & representative scale (order pC) \\
SPE rise time $\tau_r$ & representative (order ns) \\
SPE decay time $\tau_d$ & representative (order 10s ns) \\
Gain variation & channel-dependent (e.g., $\sigma_G/G \sim 10\%$) \\
Deadtime recovery model & fixed window or distributed (Sec.~2.4) \\
\hline
\end{tabular}
\label{tab:hardware_params}
\end{table}

\subsection{Liveness Function and Deadtime Injection}
Each sensor channel $i$ is associated with a binary liveness function
$L_i(t) \in \{0,1\}$, where $L_i(t)=1$ denotes a live channel and $L_i(t)=0$
indicates a non-live (deadtime) interval.
In the hybrid validation framework, deadtime windows are injected
probabilistically using a stress parameter $P_{\mathrm{dead}}$, representing the
probability of entering a deadtime state, and a window-length model chosen to
approximate typical readout recovery behavior.
Such injected non-liveness models correspond to standard deadtime abstractions
used in radiation detection and counting systems \cite{DeadtimeKnoll,NonParalyzable}.

In discrete time, the main evaluation uses a sampling rate of $f_s=\fs$,
corresponding to a sampling interval $\Delta t=1/f_s\approx16.7\,\mathrm{ns}$.
The liveness function is denoted $L_i[n]$ and applied within the DSP update stage.
Crucially, this work treats non-liveness as an explicit operational state rather
than missing data to be discarded.
In a real optical module, deadtime can be introduced by front-end effects such
as ADC clipping, finite waveform buffers, local veto/reset logic, and recovery
delays following large pulses.
The key point is that $L_i(t)=0$ does not imply the physical signal is absent;
it implies that channel $i$ is temporarily unable to deliver a valid measurement
to the trigger pipeline.
Encoding this explicitly prevents the trigger from conflating
\emph{measurement unavailability} with \emph{true silence}
\cite{IceCubeDetector,IceCubeTrigger}.

To stress-test trigger robustness in a controlled and repeatable way, we inject
deadtime synthetically after waveform construction.
For each channel, deadtime onset is governed by $P_{\mathrm{dead}}$.
When a deadtime episode occurs, its duration is drawn from either
(i) a fixed window length (deterministic recovery) or
(ii) a simple distribution about a mean length $W$ (distributed recovery),
allowing the evaluation to emulate both stable and fluctuating electronics
latency.
At the sample level, $L_i[n]$ can be applied as a lightweight gate:
\begin{equation}
\Psi_i^{\mathrm{gate}}[n] = \Psi_i[n]\,L_i[n],
\label{eq:naive_gating}
\end{equation}
where $\Psi_i[n]$ is the instantaneous per-channel observable and
$\Psi_i^{\mathrm{gate}}[n]$ is the value presented to subsequent stages.
This operation is computationally trivial and FPGA-friendly, but it is not by
itself sufficient for robust triggering.
Conventional coincidence logic implicitly treats missing hits as evidence of no
signal, so any non-live channel can break an otherwise valid coincidence chain.
By contrast, the proposed framework treats $L_i[n]$ as a first-class state:
deadtime is \emph{known} and can be handled explicitly by the liveness-aware
update law introduced next.
This enables clean comparisons under controlled parameter sweeps because the
physics-driven topology (from IceCube Open Data) is decoupled from
electronics-driven availability (via $L_i[n]$).

Throughout this work, the per-channel observable $\Psi_i[n]$ is expressed in
normalized amplitude units after pedestal subtraction and gain normalization,
such that noise fluctuations have zero mean and unit standard deviation.
The noise scale $\sigma_{\text{noise}}$ therefore denotes the standard deviation
of the noise-only distribution in these normalized $\Psi$-units.
All trigger thresholds, including the baseline threshold $\theta$ and the
proposed coherence threshold $\Gamma$, are reported in the same dimensionless
units.

Trigger thresholds for both the baseline coincidence trigger and the proposed
liveness-aware trigger were calibrated using noise-only decision windows under
live conditions ($P_{\mathrm{dead}} = 0$).
For each method, $2\times10^{7}$ independent noise realizations were generated
using the same preprocessing and decision-windowing pipeline as employed in the
signal evaluation. Lower-statistics exploratory runs were used only during method
development and yielded consistent threshold estimates.
For each trial, the maximum trigger statistic per decision window was recorded,
yielding empirical noise distributions.
Thresholds $\theta$ (baseline) and $\Gamma$ (proposed) were then selected such
that the probability of exceeding threshold matched a target false-trigger proxy
of $10^{-3}$ per decision window.
The calibrated thresholds were subsequently held fixed for all performance
evaluations under increasing deadtime probability, ensuring a consistent
false-trigger reference across trigger formulations.

Although naive gating represents non-liveness at the measurement interface, it introduces hard
discontinuities at deadtime boundaries.
From a signal-processing perspective, Eq.~(\ref{eq:naive_gating}) implicitly
assumes that information accumulated prior to deadtime becomes irrelevant once
$L_i[n]=0$.
This assumption is equivalent to enforcing an instantaneous state reset, which
is generally inconsistent with the physical persistence of optical signals and
the finite temporal extent of correlated detector responses.
As a consequence, artificial temporal gaps are introduced that bias downstream
coherence measures, fragment correlated structure across channels, and amplify
the sensitivity of trigger decisions to short-lived electronics artifacts
\cite{CoherenceDetection}.
These effects are particularly detrimental in distributed sensor networks,
where correlated activity may propagate across multiple nodes with small timing
offsets and partial overlap.
In such cases, even brief non-liveness in a subset of channels can cause a
coincidence-based trigger to misinterpret a coherent physical event as multiple
disconnected fragments.
These limitations motivate an update mechanism that preserves memory through
non-live intervals rather than collapsing state instantaneously.
The liveness model in this subsection therefore serves as an enabling mechanism:
while $L_i[n]$ encodes channel availability, the manner in which it enters the
state update determines whether temporal continuity is preserved or destroyed.
In the following subsection, we introduce a liveness-aware effective observable
based on a recursive IIR formulation that replaces hard discontinuities with a
controlled decay during non-live intervals, while remaining compatible with
real-time DSP and FPGA implementation constraints.

\subsection{Effective Observable (Recursive IIR)}
Detector non-liveness introduces abrupt discontinuities when measurements are
treated through naive gating, effectively resetting the trigger input whenever
a channel becomes unavailable.
To avoid this loss of accumulated information, we define a
continuity-preserving effective observable
$\Psi_i^{\mathrm{eff}}[n]$ using a first-order recursive Infinite Impulse Response
(IIR) formulation \cite{ProakisDSP}:
\begin{equation}
\Psi_i^{\mathrm{eff}}[n] =
k\,\Psi_i^{\mathrm{eff}}[n-1] +
(1-k)\,\Psi_i[n]\,L_i[n],
\label{eq:iir}
\end{equation}
where $L_i[n]\in\{0,1\}$ denotes channel liveness, $\Delta t$ is the sampling
interval, and the decay factor $k=\exp(-\alpha \Delta t)$ controls the balance
between temporal memory and responsiveness \cite{ProakisDSP,Yawisit:2025ipb}.
When a channel is live ($L_i[n]=1$), Eq.~(\ref{eq:iir}) reduces to an
exponentially weighted moving average, allowing the effective observable to
follow the instantaneous signal $\Psi_i[n]$ with controlled smoothing.
In contrast, during non-liveness intervals ($L_i[n]=0$), the input term vanishes
and the update is governed solely by the internal state,
\begin{equation}
\Psi_i^{\mathrm{eff}}[n] = k\,\Psi_i^{\mathrm{eff}}[n-1],
\end{equation}
leading to an exponential decay rather than an abrupt collapse.
This mechanism preserves phase and amplitude information accumulated prior to
deadtime, thereby maintaining temporal continuity across short measurement gaps.

From a signal-processing viewpoint, the recursion in
Eq.~(\ref{eq:iir}) corresponds to a causal linear time-invariant system with a
single pole at $z=k$.
For $0<k<1$, the system is bounded-input bounded-output (BIBO) stable and
implements a low-pass temporal memory with an effective time constant
$\tau\approx\alpha^{-1}$.
This guarantees that transient noise bursts or isolated measurement losses do
not induce unphysical discontinuities or unstable growth in the observable.

The formulation is inspired by the Synchromodulametry framework, which treats
coherence as a continuously evolving, hardware-native state variable rather than
a discrete coincidence condition.
By embedding temporal memory directly into the observable, the trigger operates
on a state that reflects recent correlated activity instead of instantaneous hit
availability alone.
In contrast to binary coincidence logic, which implicitly resets state during
non-liveness, the proposed observable retains a controlled persistence motivated
by the finite duration of particle-induced light emission and detector response.

At the system level, Eq.~(\ref{eq:iir}) can be interpreted as introducing a soft
persistence prior at the trigger input.
This reflects the physical expectation that correlated activity in a distributed
optical array does not vanish instantaneously when individual channels
temporarily enter deadtime.
Such behavior is common in large-scale detectors, where localized electronics
limitations coexist with globally coherent event topologies.
As a result, the effective observable provides a stable and interpretable
interface between front-end electronics behavior and higher-level trigger logic,
while remaining computationally lightweight and compatible with real-time DSP and
FPGA implementation constraints.

\subsection{Hybrid Validation Dataset}

\begin{figure}[!h]
\centering
\includegraphics[width=\diagramwidth]{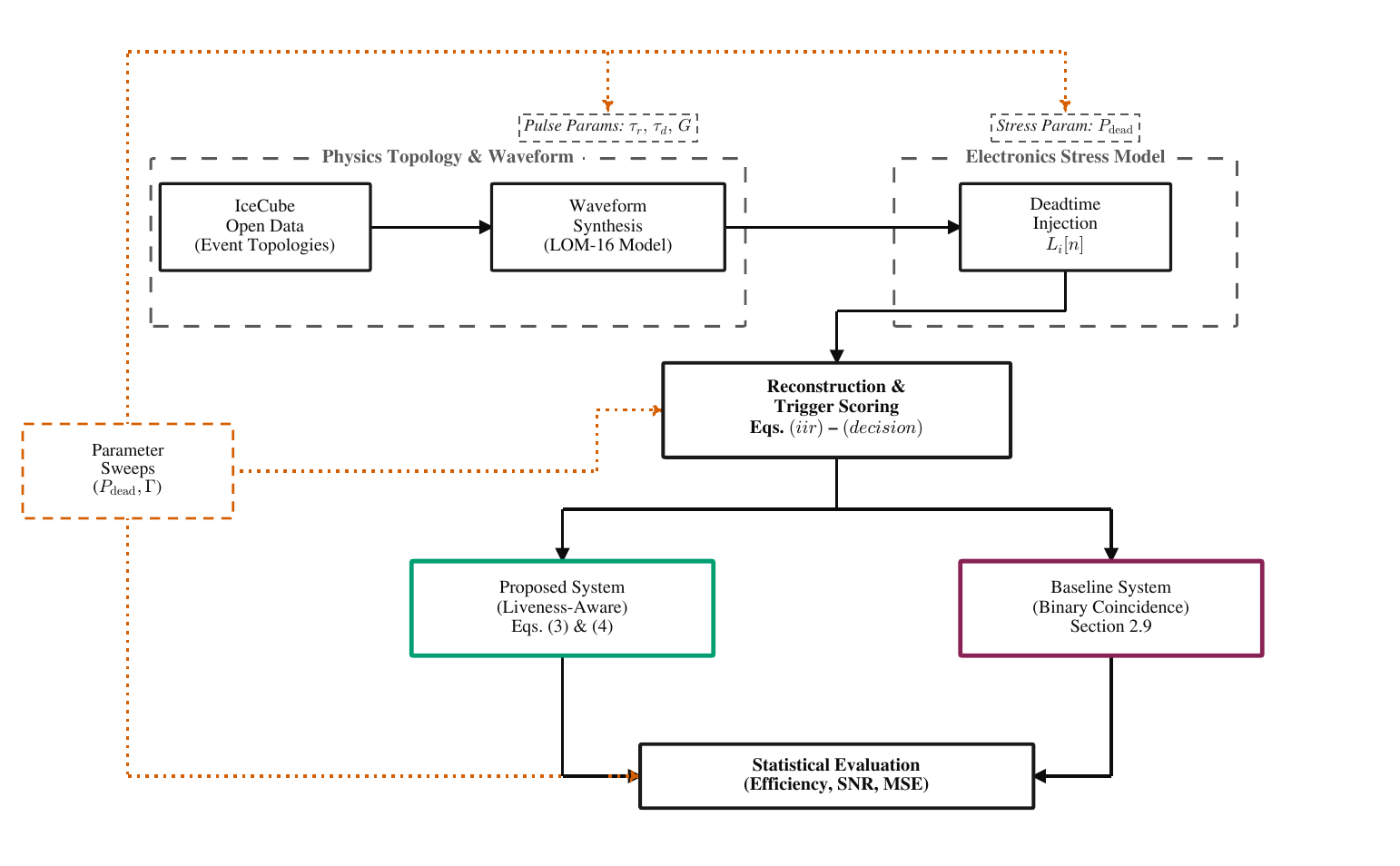}
\caption{Hybrid validation pipeline architecture.}
\label{fig:validation_pipeline}
\end{figure}

Performance validation is conducted using a hybrid dataset constructed from
(i) event topologies extracted from IceCube Open Data \cite{IceCubeOpenData},
(ii) conversion of these topologies into analog-like waveforms using the
parametric signal and readout model (Eq.~\ref{eq:spe}),
(iii) controlled deadtime injection via the liveness function $L_i[n]$, and
(iv) reconstruction and trigger scoring using
Eqs.~(\ref{eq:iir})--(\ref{eq:decision}).
This hybrid construction preserves physically meaningful temporal correlations
while enabling controlled and repeatable stress testing of trigger robustness
under electronics-induced non-liveness.

The motivation of this approach is to decouple the physics-driven structure of
an event from front-end electronics behavior.
Event topologies from IceCube Open Data encode the spatial distribution of active
channels, relative hit timing, and coarse morphological features of
neutrino- and muon-induced events, while remaining agnostic to waveform-level
digitization effects.
These topologies are held fixed across all validation runs, providing a stable
physics reference against which trigger behavior can be systematically compared.

\begin{figure}[!h]
\centering
\includegraphics[width=\wideplotwidth]{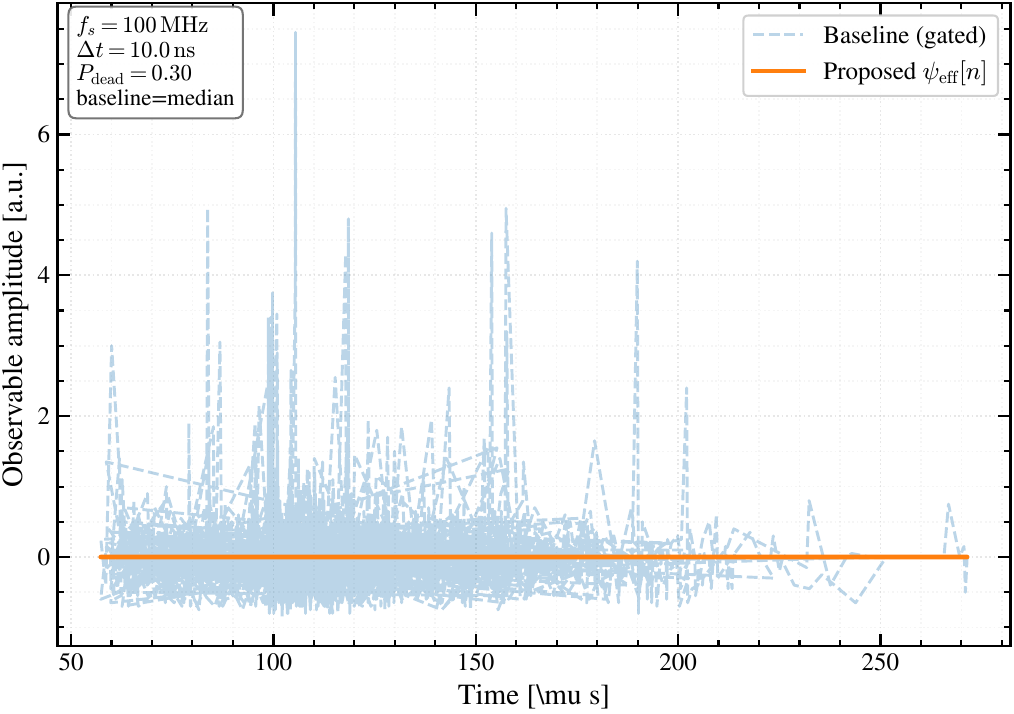}
\caption{Illustrative high-time-resolution comparison between naive gating (Baseline) and the proposed IIR-based effective observable (Proposed). This diagnostic trace was generated at $100\,\mathrm{MHz}$ ($\Delta t=10\,\mathrm{ns}$) to resolve the recursive transient; the quantitative calibration and performance sweeps use the main $60\,\mathrm{MSPS}$ configuration reported in Table~\ref{tab:trigger_settings}.}
\label{fig:iir_effect}
\end{figure}

The $100\,\mathrm{MHz}$ trace in Fig.~\ref{fig:iir_effect} is used only as a
high-time-resolution visualization of the recursive update. Unless otherwise
stated, all quantitative results, threshold calibrations, and deadtime sweeps
reported below use the $60\,\mathrm{MSPS}$ configuration.

Each fixed event topology is converted into analog-like multi-channel waveforms
using the parametric synthesis model, incorporating SPE pulse shaping, baseline
noise, gain variation, and finite ADC dynamic range.
Electronics stress is introduced through controlled deadtime injection applied
after waveform synthesis.
Deadtime windows emulate operational effects such as ADC saturation, waveform
buffer limits, and recovery delays, and are parameterized by stress variables
including deadtime probability and recovery window length.
This produces a channel-dependent liveness sequence $L_i[n]$ that encodes
measurement availability as an explicit state variable, rather than conflating
non-liveness with physical silence.

All stressed waveforms are subsequently processed by a shared reconstruction and
trigger scoring stage, which constitutes the central comparison point of the
hybrid validation pipeline.
As illustrated in Fig.~\ref{fig:validation_pipeline}, this stage applies
identical alignment, windowing, and preprocessing logic to all input waveforms
before branching into separate trigger formulations.
Specifically, both the proposed liveness-aware IIR trigger and the baseline
binary coincidence trigger receive the same reconstructed inputs, ensuring that
any downstream differences in trigger response are attributable solely to the
trigger logic itself rather than to discrepancies in upstream processing.

Within this common reconstruction framework, physics-level information and
electronics-level effects are deliberately disentangled.
Event topologies derived from IceCube Open Data determine the spatial and
temporal structure of sensor activity, while waveform synthesis translates these
topologies into realistic PMT-level signals using the parametric model.
Electronics stress is then injected explicitly via controlled deadtime windows,
which modulate the liveness function $L_i[n]$ without altering the underlying
event topology.
By holding the physical topology fixed and varying only the electronics-level
availability, the pipeline isolates trigger robustness as a function of
non-liveness rather than event morphology.

The branching into the proposed and baseline trigger paths occurs only after
this shared preprocessing stage.
In the baseline path, binary coincidence logic operates directly on thresholded
hit indicators, implicitly interpreting non-live channels as silent.
In contrast, the proposed path constructs a continuity-preserving effective
observable via the recursive IIR update, allowing coherence information to
persist across short deadtime intervals.
Both trigger outputs are subsequently passed to a unified statistical evaluation
stage, where performance metrics such as efficiency, SNR, MSE, and ROC
distributions are computed under identical parameter sweeps.

The overall structure of the hybrid validation pipeline, summarized in
Fig.~\ref{fig:validation_pipeline}, reflects a deliberate separation of concerns:
physics-level topology generation, waveform-level readout-aware synthesis, and
electronics-level stress injection are treated as distinct layers, each governed
by interpretable parameters.
This modular organization enables quantitative and reproducible assessment of
trigger behavior across a wide range of operating conditions, including varying
deadtime probability, recovery window length, and decision thresholds.

Crucially, in this formulation the effective observable in the proposed trigger
evolves continuously even in the absence of new measurements.
Short-lived non-liveness therefore results in controlled decay rather than abrupt
loss of accumulated correlation information.
This property is essential for evaluating trigger performance under realistic
electronics stress, where saturation and deadtime occur intermittently and
locally, rather than as persistent, global failures.

Stability of the recursion is guaranteed for $0 < k < 1$, and the bounded-input,
bounded-output (BIBO) property ensures that noise bursts do not lead to runaway
growth of the effective observable \cite{OppenheimDSP,IIRFilters}.

\subsection{Coherence Score and Trigger Decision}
Node-level effective observables are aggregated into a network-level trigger
variable by computing a coherence score over a sliding window.
A hardware-friendly choice is an energy-like statistic
\cite{VanTrees,CowanStats}:
\begin{equation}
\mathcal{G}[n] =
\sum_{i=1}^{N} w_i \left(\Psi_i^{\mathrm{eff}}[n]\right)^2,
\label{eq:coherence_energy}
\end{equation}
where $w_i$ are optional per-channel weights accounting for calibration or
geometric factors.
More advanced alternatives, such as covariance-based or low-rank coherence
metrics, can be incorporated within the same framework without altering the
liveness-aware update law \cite{CoherenceDetection}.

In practice, multi-node alignment can be supported by sub-nanosecond timing
distribution and synchronization schemes (e.g., White Rabbit / IEEE~1588-class
approaches), which motivate treating timing consistency as part of the trigger
measurement layer in distributed arrays \cite{WhiteRabbit,PTP1588,TimeSync,TimeDistributedDAQ,DistributedSensors}.

A trigger decision is issued when the coherence score exceeds a threshold
$\Gamma$ for at least one sample within a decision window $\mathcal{W}$:
\begin{equation}
\text{Trigger} =
\mathbf{1}\!\left(
\max_{n \in \mathcal{W}} \mathcal{G}[n] \ge \Gamma
\right),
\label{eq:decision}
\end{equation}
where $\mathbf{1}(\cdot)$ denotes the indicator function.

The energy-like coherence score in Eq.~(\ref{eq:coherence_energy}) is selected
for its simplicity, interpretability, and compatibility with fixed-point
arithmetic.
Under Gaussian noise assumptions, this statistic approximates a log-likelihood
ratio for detecting excess correlated energy across channels, and can be treated
as a windowed random-process test statistic \cite{VanTrees,CowanStats,EventPersistence}.

While higher-order statistics can improve sensitivity in some regimes, the
present choice provides a robust baseline that isolates the impact of
liveness-aware temporal continuity.
More sophisticated metrics can be incorporated in future implementations without
modifying the underlying update mechanism.

\begin{figure}[!h]
\centering
\includegraphics[width=\wideplotwidth]{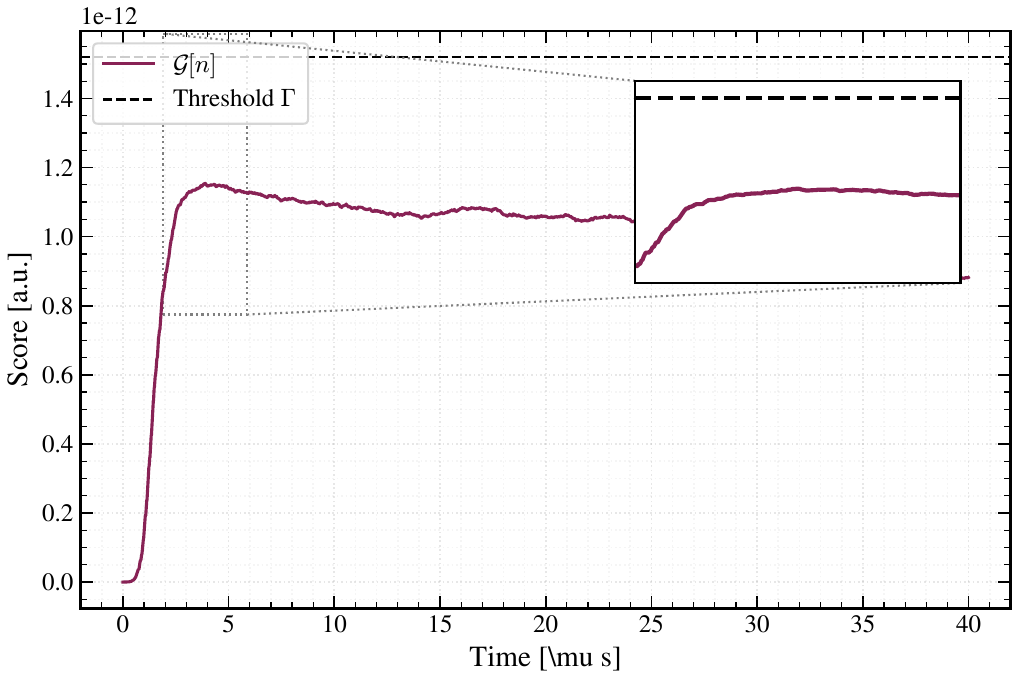}
\caption{Aggregation of node-level effective observables into a global coherence score $\mathcal{G}[n]$.}
\label{fig:coherence}
\end{figure}

A conceptual illustration of the coherence aggregation and decision logic is
shown in Fig.~\ref{fig:coherence}.
The trigger operates on a continuously evolving network-level statistic rather
than discrete coincidence events, enabling graceful degradation under partial
sensor non-liveness.

\subsection{Matrix-based coherence diagnostics}
In addition to the scalar coherence score in Eq.~(\ref{eq:coherence_energy}),
we evaluate the spatial correlation structure of the multi-channel response by
constructing a windowed coherence (correlation) matrix
$\mathbf{C}\in\mathbb{R}^{N\times N}$ from the effective observables
$\Psi_i^{\mathrm{eff}}[n]$ within each analysis window.
The eigenvalue spectrum of $\mathbf{C}$ is then used as a compact diagnostic of
intrinsic dimensionality: a small number of dominant eigenmodes indicates
low-rank coherent structure, whereas a flat spectrum is consistent with
noise-dominated or weakly correlated activity. These diagnostics provide an
interpretability layer that complements trigger-level efficiency metrics.

\subsection{Baseline Trigger for Comparison (Multiplicity Coincidence)}
\label{sec:baseline}

As a reference, we compare against a conventional multiplicity coincidence trigger.
For each channel, a binary hit indicator is formed by thresholding the instantaneous observable:
\begin{equation}
H_i[n] =
\mathbf{1}\!\left(\Psi_i[n] \ge \theta\right)\,L_i[n],
\label{eq:baseline_hit}
\end{equation}
where $\theta$ is a fixed per-channel threshold and $L_i[n]$ enforces non-liveness
($H_i[n]=0$ whenever $L_i[n]=0$).
A coincidence trigger is issued if the number of simultaneous hits exceeds a multiplicity requirement:
\begin{equation}
\text{Trigger}_{\mathrm{base}} =
\mathbf{1}\!\left(
\max_{n \in \mathcal{W}} \sum_{i=1}^{N} H_i[n] \ge M
\right),
\label{eq:baseline_mult}
\end{equation}
using the same decision window $\mathcal{W}$ as in Eq.~(\ref{eq:decision}) to ensure a fair comparison.
This baseline highlights the structural failure mode under deadtime:
temporarily unavailable channels are indistinguishable from physically silent channels,
so valid coincidence chains can fragment even when the underlying event remains coherent.

\begin{figure}[!h]
\centering
\includegraphics[width=\wideplotwidth]{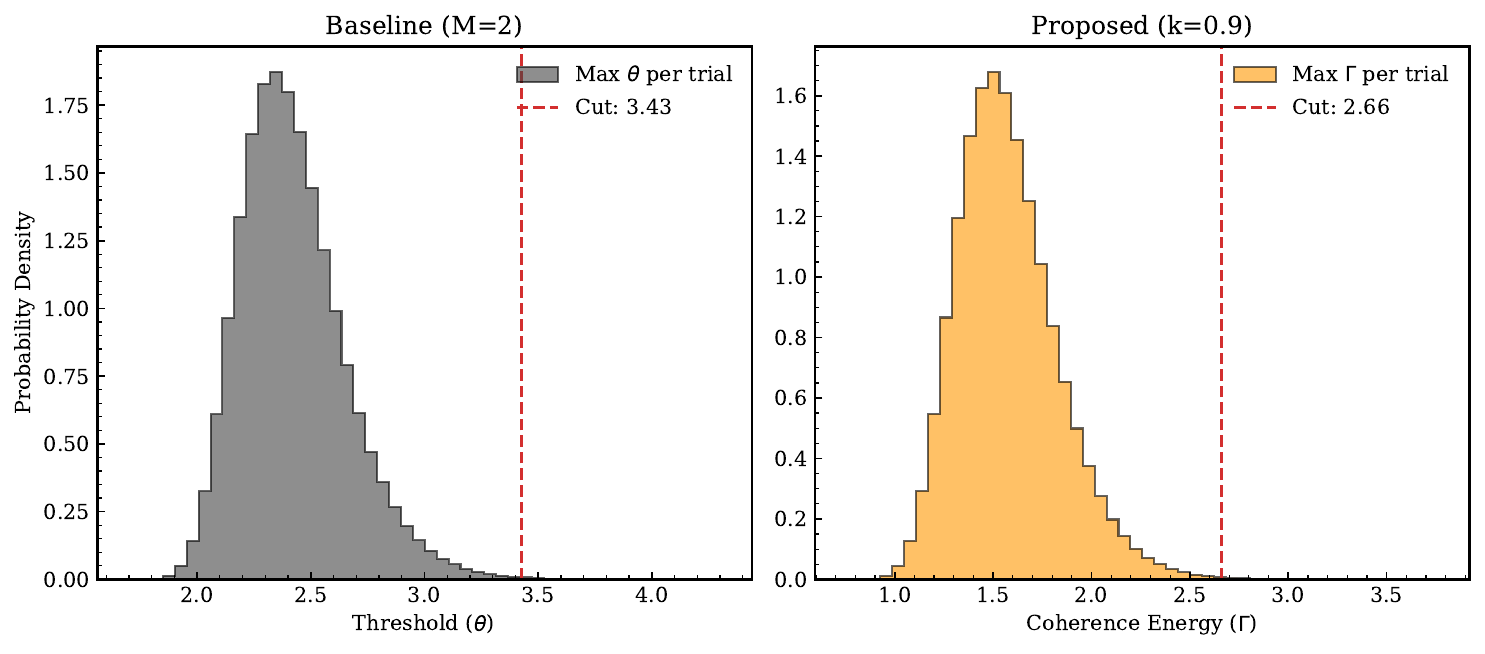}
\caption{
Noise-only distributions of the maximum trigger statistic per decision window
for the baseline coincidence trigger (left) and the proposed liveness-aware
trigger (right).
Dashed lines indicate calibrated thresholds corresponding to a target
false-trigger probability of $10^{-3}$ per decision window, obtained under live
conditions ($P_{\mathrm{dead}} = 0$).
}
\label{fig:noise_calibration}
\end{figure}

Despite its simplicity, the multiplicity coincidence trigger remains a
widely adopted baseline in large-scale optical arrays because of its
interpretability, low computational cost, and well-understood statistical
behavior under stationary noise conditions.
When channels are continuously live and noise statistics are stable,
the coincidence requirement effectively suppresses uncorrelated background
fluctuations by demanding near-simultaneous threshold crossings across
multiple sensors.

\begin{table}[!h]
\centering
\caption{Trigger configuration and threshold calibration used in this study.
Thresholds are chosen to match a target noise-only false-trigger probability of
$10^{-3}$ per decision window using $2\times10^{7}$ Monte Carlo trials (seed = 42).
Earlier exploratory calibrations using fewer trials yielded consistent values
within statistical uncertainty.}
\begin{tabular}{l c c}
\hline
Parameter & Baseline & Proposed \\
\hline
Sampling rate $f_s$ & \multicolumn{2}{c}{$60\,\mathrm{MSPS}$} \\
Channels $N$ & \multicolumn{2}{c}{16} \\
Decision window $|\mathcal{W}|$ & \multicolumn{2}{c}{$1.5\,\mu\mathrm{s}$} \\
Noise-only calibration & \multicolumn{2}{c}{$2\times10^{7}$ trials, target $10^{-3}$} \\
Threshold & $\theta = 3.428$ & $\Gamma = 2.659$ \\
Multiplicity / Score & $M=2$ & $\mathcal{G}[n]$ (Eq.~\ref{eq:coherence_energy}) \\
IIR persistence ($k$) & -- & 0.90 \\
Achieved false-proxy & $1.0\times10^{-3}$ & $1.0\times10^{-3}$ \\
\hline
\end{tabular}
\label{tab:trigger_settings}
\end{table}

For this reason, coincidence logic provides a meaningful reference point
against which alternative trigger formulations can be evaluated, particularly
when thresholds and decision windows are calibrated to identical false-trigger
targets.

However, the coincidence formulation embeds a strong implicit modeling
assumption: channel availability is binary and perfectly correlated with
physical silence.
The liveness mask $L_i[n]$ enforces this assumption explicitly by forcing
$H_i[n]=0$ whenever a channel is non-live, regardless of the underlying
physical signal state.
As a consequence, coincidence logic cannot distinguish between the absence of
a signal and the absence of a measurement.
This structural ambiguity becomes critical under realistic operating
conditions, where localized deadtime, saturation recovery, or buffering delays
occur intermittently and asynchronously across channels.
Even short non-liveness intervals can therefore break otherwise valid
coincidence chains, leading to a disproportionate loss of trigger efficiency
that is not recoverable through threshold retuning alone.

Importantly, the baseline trigger defined in
Eqs.~(\ref{eq:baseline_hit})--(\ref{eq:baseline_mult}) is intentionally kept
minimal and unmodified throughout this study.
All thresholds and decision windows are calibrated using the same noise-only
procedure as the proposed trigger and are held fixed across deadtime sweeps.
This ensures that observed performance differences arise from the structural
treatment of non-liveness rather than from parameter optimization or algorithmic
complexity.
Under this controlled comparison, the baseline coincidence trigger serves as a
clear illustration of how electronics-induced measurement loss is propagated
directly into trigger-level decision failures when liveness is not modeled as an
explicit state variable.

\clearpage
\section{Results}
\label{sec:results}

\subsection{Qualitative Reconstruction Under Non-Liveness}

\begin{figure}[!h]
\centering
\includegraphics[width=\wideplotwidth]{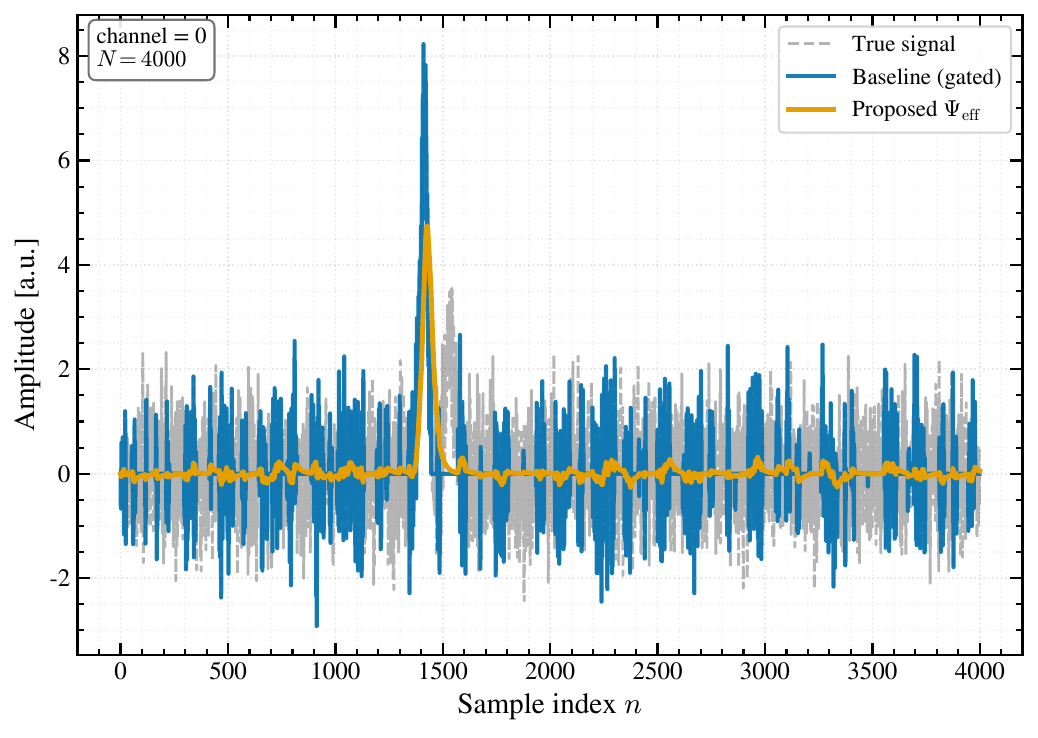}
\caption{Qualitative comparison of signal reconstruction under deadtime.}
\label{fig:recon}
\end{figure}

Figure~\ref{fig:recon} presents a representative reconstruction example
illustrating the qualitative behavior of the proposed liveness-aware trigger
under injected deadtime.
In the baseline coincidence-driven pipeline, the gated signal
(Eq.~\ref{eq:naive_gating}) collapses whenever one or more channels enter a
non-live state.
As a consequence, temporal continuity is lost and network-level correlation
structure becomes fragmented.

In contrast, the proposed effective observable $\Psi_i^{\mathrm{eff}}[n]$
(Eq.~\ref{eq:iir}) exhibits a smooth, continuous decay during non-liveness
intervals and recovers naturally once liveness is restored.
This behavior preserves phase- and amplitude-like continuity across short
deadtime gaps, preventing artificial segmentation of correlated activity in the
multi-channel response.

Figure~\ref{fig:recon} visualizes this qualitative difference by comparing the
time evolution of the gated signal and the proposed effective observable under
identical deadtime conditions.
The baseline approach exhibits abrupt signal collapse during non-liveness,
whereas the proposed IIR effective observable maintains temporal continuity
through controlled decay and recovery.
The continuity preserved by $\Psi_i^{\mathrm{eff}}[n]$ enables downstream
coherence metrics to remain sensitive to correlated activity even when
individual sensor streams temporarily fail to report valid measurements.


\subsection{Trigger Efficiency Versus Deadtime}
Trigger efficiency is evaluated as a function of the deadtime probability
$P_{\mathrm{dead}}$ while holding all other parameters fixed.
Efficiency is defined as the fraction of injected signal events whose coherence
score exceeds the trigger threshold $\Gamma$ at least once within the decision
window $\mathcal{W}$ (Eq.~\ref{eq:decision}).

\begin{figure}[!h]
\centering
\includegraphics[width=\plotwidth]{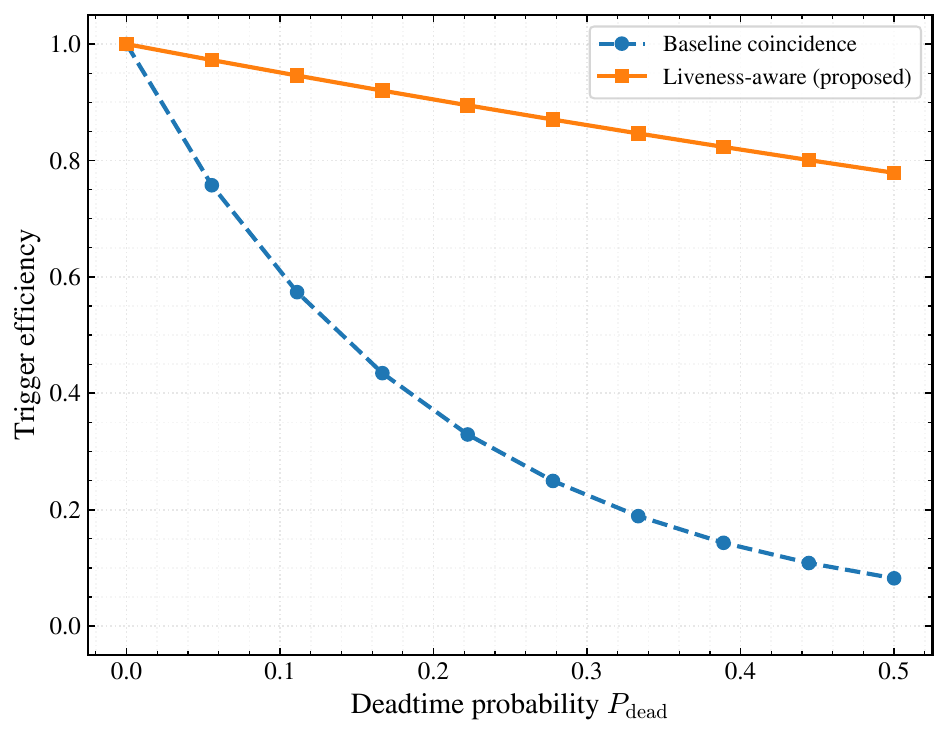}
\caption{Trigger efficiency as a function of deadtime probability $P_{\mathrm{dead}}$.}
\label{fig:eff_deadtime}
\end{figure}

Figure~\ref{fig:eff_deadtime} compares the proposed liveness-aware trigger to a
baseline coincidence-based trigger.
At low deadtime probability, both methods achieve comparable efficiency,
indicating that the proposed approach does not sacrifice nominal performance.
As $P_{\mathrm{dead}}$ increases, the baseline trigger exhibits a rapid
degradation in efficiency due to fragmentation of coincidence chains.
In contrast, the liveness-aware trigger sustains higher efficiency over a broad
range of moderate deadtime levels.

At sufficiently large $P_{\mathrm{dead}}$, efficiency eventually degrades for
both methods, reflecting the fundamental loss of information when a substantial
fraction of channels is non-live.
However, the onset of efficiency loss occurs at systematically higher
$P_{\mathrm{dead}}$ for the proposed trigger, demonstrating increased robustness
to intermittent measurement availability.

The efficiency curves in Fig.~\ref{fig:eff_deadtime} are averaged over multiple
independent realizations of deadtime injection to suppress statistical
fluctuations.
Error bars (where shown) indicate one standard deviation across trials, ensuring
that observed performance trends are not driven by specific deadtime
realizations.

Overall, the proposed trigger delays the onset of efficiency loss by preserving
coherence across short non-liveness intervals, whereas coincidence logic
fragments as soon as liveness becomes intermittent.


\subsection{Coherence Structure and Low-Rank Modes}

\begin{figure}[!h]
\centering
\begin{minipage}[t]{0.48\linewidth}
  \centering
  \includegraphics[width=\linewidth]{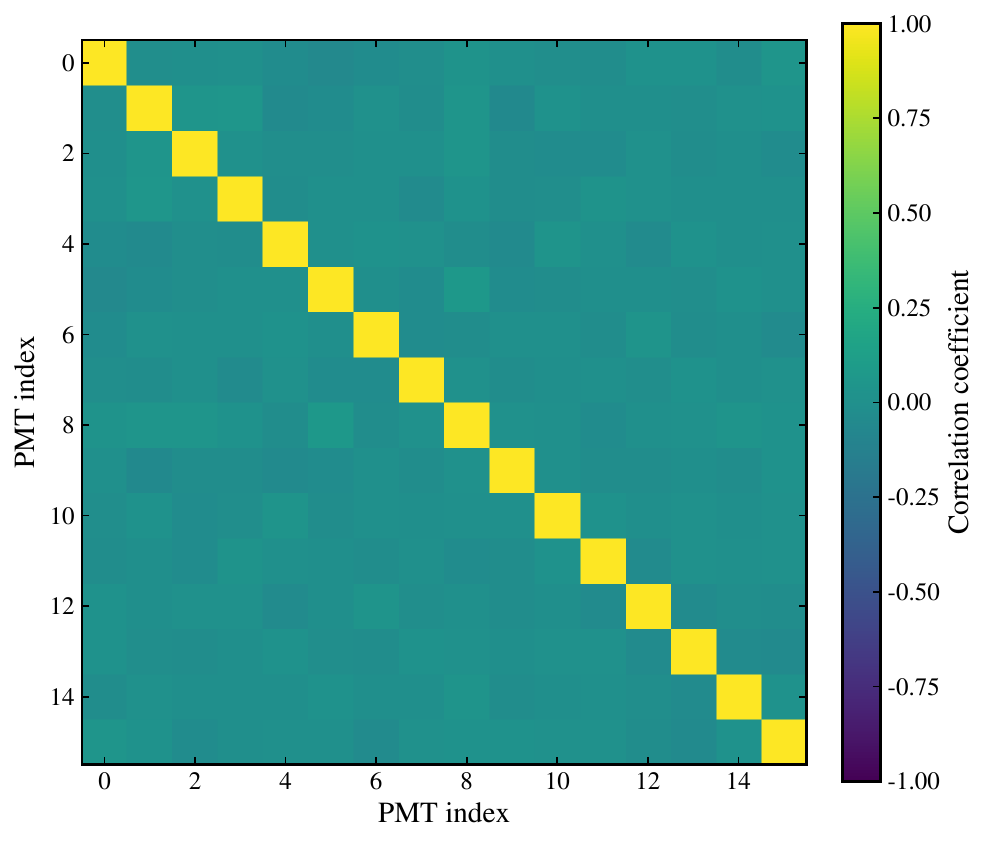}\\[-0.4ex]
  \small (a) Coherence matrix $\mathbf{C}$
\end{minipage}\hfill
\begin{minipage}[t]{0.48\linewidth}
  \centering
  \includegraphics[width=\linewidth]{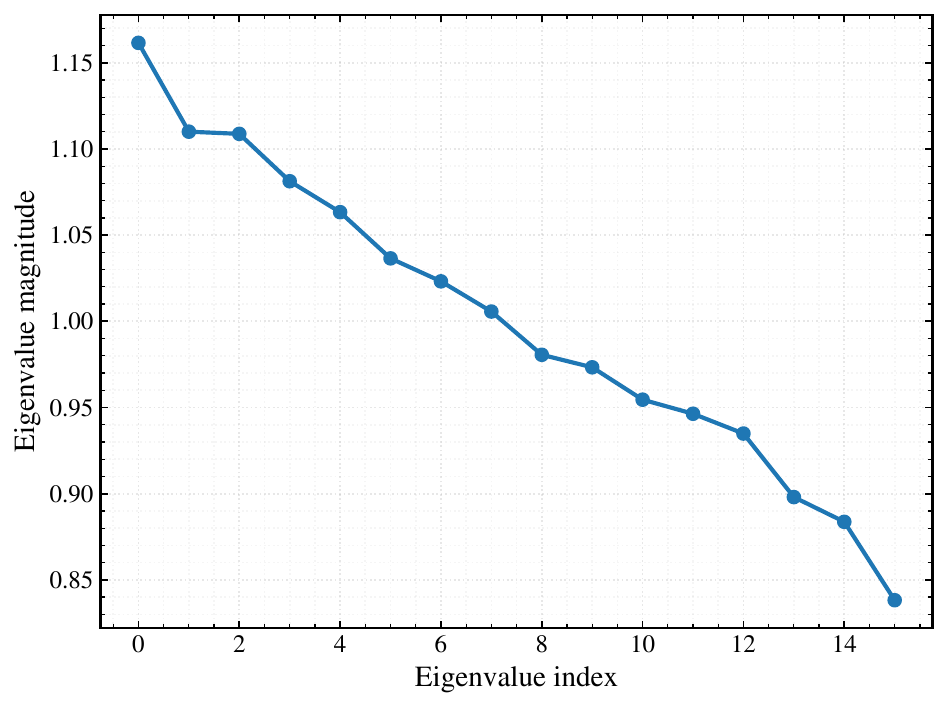}\\[-0.4ex]
  \small (b) Eigenvalue spectrum
\end{minipage}
\caption{Matrix-based coherence diagnostics: (a) the windowed inter-channel
correlation matrix computed from the effective observables and (b) its
corresponding eigenvalue spectrum. Dominant eigenmodes indicate coherent,
low-rank structure in the multi-channel response.}
\label{fig:coherence_diagnostics}
\end{figure}

Beyond scalar performance curves, we examine the correlation structure of the
multi-channel response to understand \emph{why} liveness-aware triggering remains
robust under increasing non-liveness.
Figure~\ref{fig:coherence_diagnostics}(a) shows a representative coherence (correlation)
matrix computed from the effective observables within a fixed analysis window.
The matrix exhibits clear non-random structure, with elevated correlations
concentrated among subsets of channels, indicating that the response is not a
collection of independent hits but contains coherent multi-channel patterns.

To quantify this structure, Fig.~\ref{fig:coherence_diagnostics}(b) presents the eigenvalue
spectrum of the coherence matrix.
The spectrum is strongly non-flat, with a small number of dominant eigenvalues
followed by a decaying tail.
This behavior is consistent with a low-rank representation in which a limited
set of coherent modes captures a large fraction of the correlated activity.
In contrast, a noise-dominated response would produce a comparatively flatter
spectrum.
The presence of dominant modes supports the key design premise of this work:
trigger logic can benefit from operating on continuity-preserving coherent state
variables rather than instantaneous binary coincidence statements.


\subsection{Signal-to-Noise Ratio Enhancement}
To quantify the impact of liveness-aware continuity on signal fidelity, we
evaluate the effective signal-to-noise ratio (SNR) of the reconstructed trigger
observable.
SNR is defined as the ratio between the mean squared amplitude of
signal-dominated intervals and the variance observed in noise-only regions.

\begin{figure}[tbp]
\centering
\begin{minipage}[t]{0.48\linewidth}
  \centering
  \includegraphics[width=\linewidth]{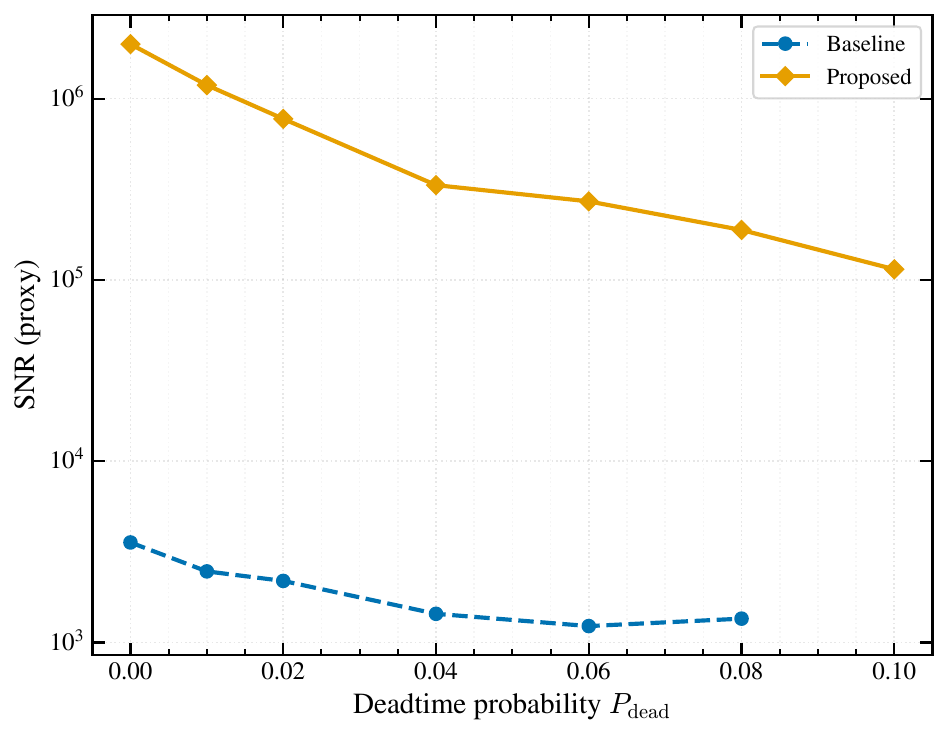}\\[-0.4ex]
  \small (a) Signal-to-noise ratio
\end{minipage}\hfill
\begin{minipage}[t]{0.48\linewidth}
  \centering
  \includegraphics[width=\linewidth]{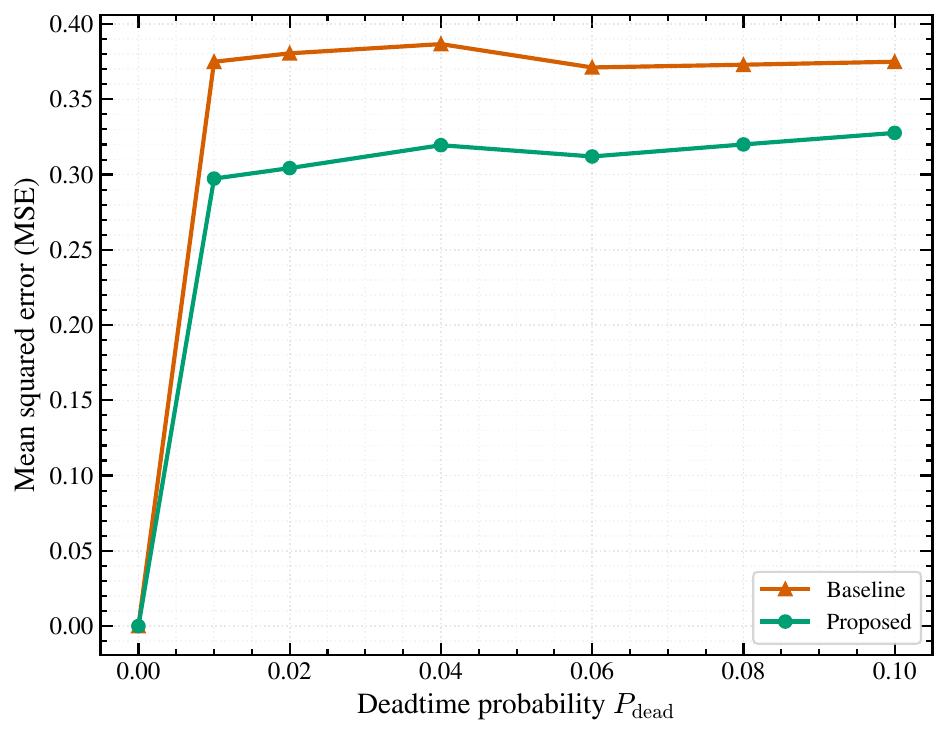}\\[-0.4ex]
  \small (b) Reconstruction error
\end{minipage}
\caption{Signal fidelity under increasing deadtime: (a) effective
signal-to-noise ratio and (b) mean squared reconstruction error evaluated over
deadtime intervals.}
\label{fig:snr_error}
\end{figure}

Figure~\ref{fig:snr_error}(a) compares the SNR achieved by the proposed trigger and the
baseline coincidence-based approach as a function of deadtime probability.
While both methods exhibit similar SNR under low-deadtime conditions, the
baseline SNR degrades rapidly as deadtime increases due to collapse of coherent
structure induced by hard gating.
In contrast, the proposed method maintains substantially higher SNR over a wide
range of operating conditions, indicating improved retention of correlated
information used for downstream decision-making.
Note that the vertical axis is logarithmic, emphasizing the separation between
methods at moderate-to-high $P_{\mathrm{dead}}$.


\subsection{Reconstruction Error Under Deadtime}
To assess reconstruction fidelity during non-liveness intervals, we compute the
difference between the ground-truth injected signal and the reconstructed
effective observable.
Error is quantified using the mean squared error (MSE) evaluated over deadtime
windows only.

Figure~\ref{fig:snr_error}(b) shows that the baseline trigger exhibits large
reconstruction error during deadtime, reflecting loss of signal information
under hard gating.
The proposed method suppresses this error by preserving temporal memory through
the recursive update law.
These results indicate that the liveness-aware effective observable maintains a
closer approximation to the underlying signal during measurement gaps.


\subsection{Robustness to Saturation and Clipping Scenarios}

\begin{figure}[!h]
\centering
\begin{minipage}[t]{0.48\linewidth}
  \centering
  \includegraphics[width=\linewidth]{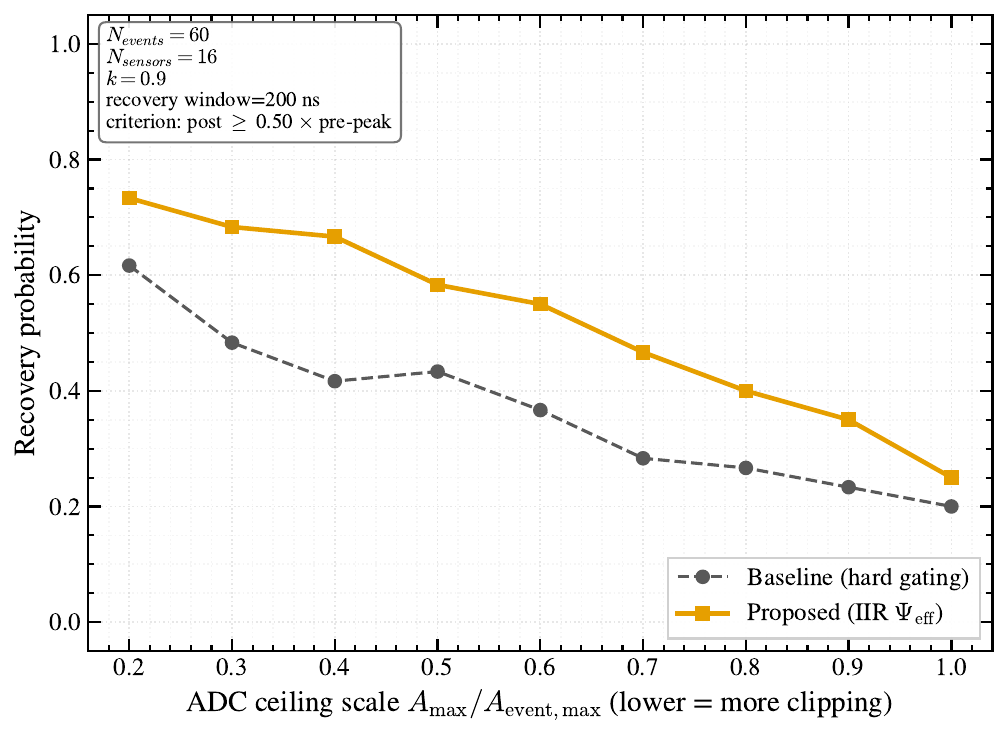}\\[-0.4ex]
  \small (a) Saturation recovery
\end{minipage}\hfill
\begin{minipage}[t]{0.48\linewidth}
  \centering
  \includegraphics[width=\linewidth]{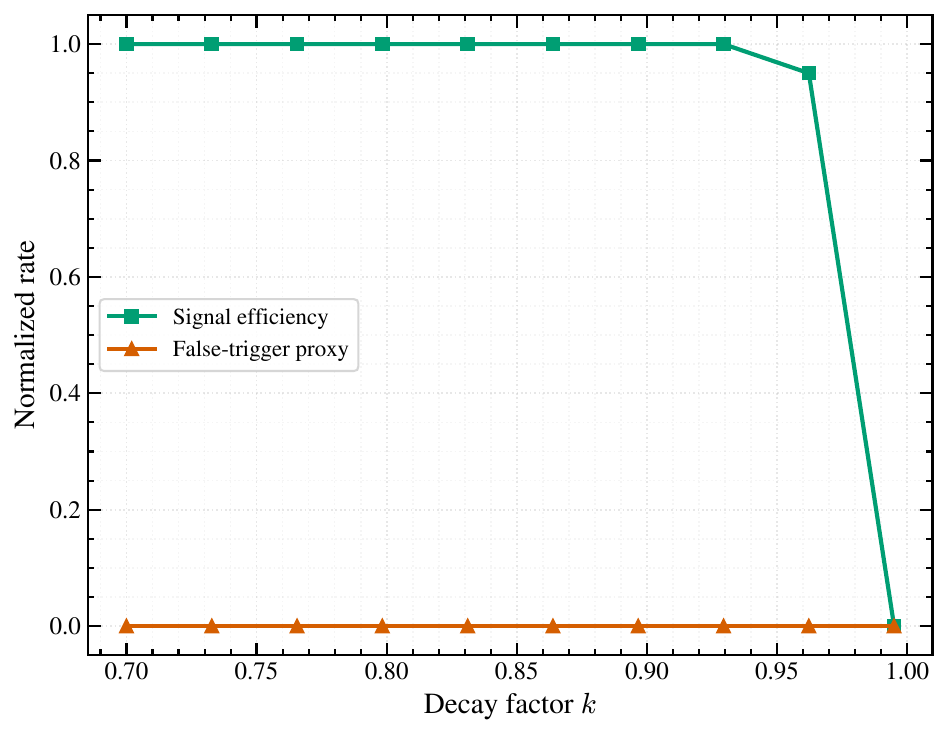}\\[-0.4ex]
  \small (b) Decay-factor sensitivity
\end{minipage}
\caption{Robustness and parameter sensitivity of the proposed trigger:
(a) recovery probability under increasing saturation/clipping severity and
(b) trigger efficiency and false-trigger proxy as functions of the IIR decay
factor $k$.}
\label{fig:robustness}
\end{figure}

To emulate realistic front-end electronics constraints, saturation and clipping
severity are varied by adjusting the maximum allowed pulse amplitude and the
frequency of ADC overflow events in the signal model.
Recovery rate is defined as the fraction of events whose coherence score returns
above threshold following a saturation-induced non-liveness interval.

As shown in Fig.~\ref{fig:robustness}(a), the baseline trigger is highly sensitive
to saturation severity, with recovery probability decreasing sharply as clipping
frequency increases.
This behavior arises because saturated channels are interpreted as inactive,
leading to persistent breaks in coincidence structure.
The proposed liveness-aware trigger remains comparatively stable across the same
parameter range.
Because the effective observable does not rely on instantaneous hit reporting
from every channel, partial saturation does not immediately suppress network
coherence.
Instead, coherence decays gradually and recovers smoothly once channels regain
liveness.
These results indicate that the proposed approach is resilient to realistic
electronics limitations that are unavoidable in large-scale optical arrays.


\subsection{Parameter Sensitivity and Stability of the Decay Factor \texorpdfstring{$k$}{k}}
The decay factor $k$ governs the temporal memory of the effective observable and
therefore represents a key design parameter.
Small values of $k$ correspond to short memory and rapid decay, while larger
values preserve coherence over longer non-liveness intervals.

We define a false-trigger proxy as the probability that the trigger statistic
crosses threshold in noise-only windows constructed from baseline pedestal plus
SPE-like noise fluctuations, with no injected physics signal.
This proxy is not a full detector background model, but provides a consistent
stability check when sweeping the persistence parameter $k$.

Figure~\ref{fig:robustness}(b) shows trigger efficiency and a false-trigger proxy
metric as functions of $k$.
For small $k$, performance approaches that of instantaneous gating, and the
benefit of liveness-aware continuity is diminished.
For excessively large $k$, coherence persists for long durations, increasing
susceptibility to noise accumulation and spurious triggers.

An intermediate operating region is observed in which efficiency is maximized
while stability is maintained.
This region defines a practical range of $k$ values suitable for deployment and
demonstrates that the proposed trigger can be tuned systematically rather than
heuristically.

Beyond the optimal operating region, the observed degradation defines a clear
boundary for stable operation.
This boundary provides a quantitative guideline for selecting decay parameters
based on expected deadtime characteristics and noise environment.
Such tunability is essential for adapting the trigger to different detector
configurations without redesigning the underlying architecture.

The signal efficiency remains high over a wide range ($k \in [0.70, 0.95]$),
while the false-trigger proxy remains negligible until $k$ approaches unity,
where persistence can lead to spurious triggers. The wide stable plateau
indicates robust operation without the need for precise fine-tuning.

\clearpage
\section{Discussion and Conclusion}
\label{sec:discussion-conclusion}

\subsection{Interpretation and Comparison with Conventional Triggers}
A central limitation of coincidence-based triggers under deadtime is structural:
coincidence logic reduces each sensor stream to a binary hit/no-hit statement and
therefore implicitly assumes that the absence of a hit is informative evidence
against the presence of a correlated physical signal.
When a subset of channels becomes non-live, a conventional trigger cannot
distinguish between (i) a physically quiet channel and (ii) an unavailable
measurement channel. As a result, valid multi-channel correlations fragment
into disconnected coincidence chains, and trigger evidence collapses even when
the underlying event remains coherent in space and time.

The matrix diagnostics provide a complementary explanation for the efficiency
trends. The coherence matrix reveals structured inter-channel correlations, and
its eigenvalue spectrum indicates that the response is effectively low-rank,
dominated by a small number of coherent modes. Coincidence logic discards this
structure by reducing each channel to a binary hit/no-hit statement and by
treating non-liveness as physical silence. The proposed trigger, in contrast,
preserves a continuous state representation through deadtime, allowing these
coherent modes to remain detectable even under intermittent measurement loss.

The proposed trigger addresses this failure mode by treating liveness as an
explicit operational state and by separating \emph{measurement construction} from
\emph{decision logic}. At the measurement layer, each channel maintains a
continuity-preserving effective observable via the recursive update in
Eq.~(\ref{eq:iir}). During non-liveness ($L_i[n]=0$), the observable does not
drop discontinuously to zero, but decays smoothly according to the internal state
term $k\,\Psi_i^{\mathrm{eff}}[n-1]$. This mechanism preserves a bounded memory of
recent activity and prevents the abrupt ``information reset'' that is
characteristic of naive gating or binary hit formation. In turn, network-level
aggregation (Eq.~\ref{eq:coherence_energy}) operates on a state variable that
remains well-defined even when individual channels temporarily fail to report
new samples. Practically, this converts the trigger problem from a fragile
all-or-nothing coincidence requirement into a continuous coherence tracking task.

The qualitative behavior is visible in Fig.~\ref{fig:iir_effect} (and the
reconstruction example in Fig.~\ref{fig:recon}): naive gating introduces hard
gaps that artificially de-phase the multi-channel structure, whereas the IIR
state maintains continuity across short deadtime windows. Quantitatively, the
same mechanism explains why efficiency in Fig.~\ref{fig:eff_deadtime} degrades
more slowly for the proposed trigger as $P_{\mathrm{dead}}$ increases: the
trigger does not require every participating channel to contribute instantaneous
hits at the same time, but instead accumulates coherent energy through a state
that is robust to intermittent availability. The SNR and MSE trends
(Fig.~\ref{fig:snr_error}) further support this interpretation:
when deadtime increases, baseline coincidence logic loses correlation structure,
which simultaneously reduces detection probability and degrades the quality of
the trigger-level observable. In contrast, the liveness-aware recursion retains
a physically motivated continuity prior, correlated activity does not vanish
instantaneously, and therefore better preserves usable information for the
decision layer.

An important point is that the observed improvement is not achieved by relaxing
physics requirements or by artificially inflating the signal, but by preventing
electronics-induced unavailability from being misinterpreted as physics-level
silence. The hybrid validation pipeline (Fig.~\ref{fig:validation_pipeline})
was designed specifically to isolate this effect: event topologies from IceCube
Open Data are held fixed, while electronics stress is varied through controlled
deadtime and saturation behavior. Under this construction, differences in
trigger performance can be attributed directly to how the trigger handles
non-liveness, rather than to changes in event morphology. This provides a clean
explanation for why the proposed approach is particularly beneficial in
next-generation deployments, where heterogeneous channel response and
intermittent measurement loss are expected operating conditions rather than
exceptional cases.

\subsection{Engineering Implications and FPGA Readiness}
From an implementation standpoint, the proposed trigger is intentionally aligned
with real-time constraints. The per-channel recursion in Eq.~(\ref{eq:iir}) has
$\mathcal{O}(1)$ complexity per sample and requires only minimal state (one
stored value per channel). In fixed-point hardware, the update can be
implemented as a single multiply-accumulate plus lightweight liveness gating.
Similarly, the coherence statistic in Eq.~(\ref{eq:coherence_energy}) is an
energy-like accumulation that maps naturally onto DSP slices and pipelined adder
trees. Because neither stage requires matrix operations, iterative solvers, or
global buffering, the latency is predictable and amenable to timing closure.

The explicit separation between measurement and decision layers also has
practical engineering value. Thresholds, window lengths, and trigger logic can
be tuned for bandwidth and false-rate targets without destabilizing the
measurement state. Conversely, the measurement layer can be optimized for signal
fidelity (e.g., quantization strategy, per-channel weighting, or time-alignment
precision) without modifying decision semantics. This modularity is advantageous
for staged deployment: a firmware prototype can first validate the per-channel
recursion and coherence accumulation on streaming data, then incrementally
incorporate calibration weights $w_i$, geometry-aware alignment, and more refined
decision logic. In addition, the tunable parameter $k$ provides an explicit
control knob for balancing persistence and responsiveness, enabling the trigger
to be adapted to different detector configurations and expected deadtime profiles
without changing the underlying hardware architecture.

\subsection{Limitations and Future Work}
This work presents a proof-of-design validated through a hybrid simulation
framework. Several limitations motivate clear next steps. First, the coherence
score used here is intentionally simple (energy-like) to isolate the effect of
liveness-aware continuity. While this choice is robust and hardware-friendly, it
does not exhaust the full space of coherence metrics; covariance-aware,
phase-sensitive, or low-rank formulations could increase sensitivity in specific
regimes and will be investigated in future iterations. Second, deadtime and
saturation behavior are injected through a controlled model rather than derived
from a full electronics simulation. Although the injected liveness windows are
physically motivated (ADC clipping, buffer limits, recovery delays), future work
should incorporate richer front-end behavior, including rate-dependent recovery,
baseline shifts, dual-gain paths, and channel-specific firmware logic.

Third, the present validation uses a hybrid dataset that preserves realistic
topology while enabling controlled sweeps, but it remains a simulation-layer
evaluation. A hardware-in-the-loop test (streaming digitized waveforms through
an FPGA prototype) would strengthen the link to deployment by quantifying
fixed-point effects, resource utilization, and end-to-end latency under realistic
throughput. In addition, broader evaluation across multiple event classes and
background conditions is required to fully characterize trade-offs in efficiency
and false-trigger behavior. Finally, while the present study uses a
hardware-informed signal model to emulate realistic saturation and recovery
effects, the method is designed to be portable; future work will apply the same
pipeline to alternative front-end configurations and digitizer settings to test
generality.

\subsection{Conclusion}
This paper presented the design and implementation of a liveness-aware trigger
system for distributed optical arrays, motivated by the unavoidable presence of
deadtime and electronics-induced measurement loss in large-scale observatories.
By introducing an explicit liveness state and a continuity-preserving IIR
effective observable, the proposed trigger maintains a coherent internal
representation across non-live intervals and avoids the structural fragility of
binary coincidence logic. Using a hybrid validation pipeline that couples real
IceCube Open Data topologies with a hardware-informed pulse and readout model and
controlled deadtime stress testing, the results demonstrate improved event
recovery and signal fidelity under high-deadtime conditions while maintaining
nominal performance in low-deadtime regimes. The computational simplicity and
deterministic structure of the approach support practical implementation in
real-time FPGA-based trigger systems, providing a scalable foundation for
next-generation DAQ architectures in high-energy neutrino and multimessenger
observatories.

\begin{acknowledgement}
The authors would like to thank advisors, collaborators, and colleagues for
helpful discussions and technical feedback related to real-time triggering,
detector readout, and firmware-oriented signal processing, which contributed to
the development and refinement of this work. The authors gratefully acknowledge the IceCube Collaboration for making IceCube Open Data publicly available, which was used as a physics-motivated basis for the hybrid validation pipeline employed in this study. Any interpretations,
conclusions, or remaining errors are the responsibility of the authors and do
not necessarily reflect the views of the IceCube Collaboration. The first author acknowledges financial and scholarship support received during the broader research and training period associated with this work, including institutional and external scholarship programs.
\end{acknowledgement}

\clearpage


\begin{thebibliography}{99}

\bibitem{IceCubeScience}
M.~G.~Aartsen \emph{et al.} (IceCube Collaboration),
``Evidence for High-Energy Extraterrestrial Neutrinos at the IceCube Detector,''
\emph{Science}, vol.~342, no.~6161, 2013.
\href{https://doi.org/10.1126/science.1242856}{doi:10.1126/science.1242856}

\bibitem{IceCubeDetector}
M.~G.~Aartsen \emph{et al.} (IceCube Collaboration),
``The IceCube Neutrino Observatory: Instrumentation and Online Systems,''
\emph{Journal of Instrumentation}, vol.~12, P03012, 2017.
\href{https://doi.org/10.1088/1748-0221/12/03/P03012}{doi:10.1088/1748-0221/12/03/P03012}

\bibitem{IceCubeTrigger}
M.~G.~Aartsen \emph{et al.} (IceCube Collaboration),
``Online reconstruction and filtering in IceCube,''
\emph{Nucl. Instrum. Methods A}, vol.~736, pp.~143--149, 2014.
\href{https://doi.org/10.1016/j.nima.2013.10.074}{doi:10.1016/j.nima.2013.10.074}

\bibitem{Gen2Overview}
M.~G.~Aartsen \emph{et al.} (IceCube-Gen2 Collaboration),
``IceCube-Gen2: A vision for the future of neutrino astronomy,''
\emph{Journal of Physics G}, vol.~48, 060501, 2021.
\href{https://doi.org/10.1088/1361-6471/abeb9d}{doi:10.1088/1361-6471/abeb9d}

\bibitem{IceCubeOpenData}
IceCube Collaboration,
``IceCube Open Data,''
\href{https://icecube.wisc.edu/science/data/}{https://icecube.wisc.edu/science/data/}

\bibitem{KM3NeT}
S.~Adrian-Martinez \emph{et al.} (KM3NeT Collaboration),
``Letter of intent for KM3NeT 2.0,''
\emph{Journal of Physics G}, vol.~43, 084001, 2016.
\href{https://doi.org/10.1088/0954-3899/43/8/084001}{doi:10.1088/0954-3899/43/8/084001}

\bibitem{BaikalGVD}
A.~Avrorin \emph{et al.},
``The Baikal-GVD neutrino telescope,''
\emph{Nucl. Instrum. Methods A}, vol.~742, pp.~82--88, 2014.
\href{https://doi.org/10.1016/j.nima.2013.12.023}{doi:10.1016/j.nima.2013.12.023}

\bibitem{MultiMessenger}
B.~P.~Abbott \emph{et al.},
``Multimessenger observations of a binary neutron star merger,''
\emph{Astrophysical Journal Letters}, vol.~848, L12, 2017.
\href{https://doi.org/10.3847/2041-8213/aa91c9}{doi:10.3847/2041-8213/aa91c9}

\bibitem{Yawisit:2025ipb}
T.~Yawisit,
\emph{Synchromodulametry: A Hardware-First, Metric-Aware Measurement Interface for Multimessenger Coherence},
\href{https://arxiv.org/abs/2512.20678}{arXiv:2512.20678} [physics.ins-det] (2025),
\href{https://doi.org/10.48550/arXiv.2512.20678}{doi:10.48550/arXiv.2512.20678}.

\bibitem{DAQReview}
P.~C.~Bhat,
``Trigger and data acquisition systems,''
\emph{Annual Review of Nuclear and Particle Science},
vol.~41, pp.~115--162, 1991.
\href{https://doi.org/10.1146/annurev.ns.41.120191.000555}{doi:10.1146/annurev.ns.41.120191.000555}

\bibitem{DeadtimeKnoll}
G.~F.~Knoll,
\emph{Radiation Detection and Measurement},
4th ed., Wiley, 2010.

\bibitem{NonParalyzable}
J.~C.~Davis,
``Dead-time corrections for counting systems,''
\emph{Nucl. Instrum. Methods},
vol.~18, pp.~1--7, 1962.
\href{https://doi.org/10.1016/0029-554X(62)90127-6}{doi:10.1016/0029-554X(62)90127-6}

\bibitem{CMSDAQ}
S.~Chatrchyan \emph{et al.} (CMS Collaboration),
``The CMS trigger system,''
\emph{Journal of Instrumentation},
vol.~12, P01020, 2017.
\href{https://doi.org/10.1088/1748-0221/12/01/P01020}{doi:10.1088/1748-0221/12/01/P01020}

\bibitem{CowanStats}
G.~Cowan,
\emph{Statistical Data Analysis},
Oxford University Press, 1998.

\bibitem{VanTrees}
H.~L.~Van Trees,
\emph{Detection, Estimation, and Modulation Theory},
Wiley, 2001.

\bibitem{EventPersistence}
A.~Papoulis,
\emph{Probability, Random Variables, and Stochastic Processes},
McGraw-Hill, 2002.

\bibitem{MatchedFiltering}
B.~P.~Abbott \emph{et al.} (LIGO Scientific Collaboration),
``Search for gravitational waves using matched filtering,''
\emph{Physical Review D},
vol.~69, 122001, 2004.
\href{https://doi.org/10.1103/PhysRevD.69.122001}{doi:10.1103/PhysRevD.69.122001}

\bibitem{Kalman}
R.~E.~Kalman,
``A new approach to linear filtering and prediction problems,''
\emph{ASME Journal of Basic Engineering},
vol.~82, pp.~35--45, 1960.
\href{https://doi.org/10.1115/1.3662552}{doi:10.1115/1.3662552}

\bibitem{StateSpaceDetection}
L.~Ljung,
\emph{System Identification: Theory for the User},
2nd ed., Prentice Hall, 1999.

\bibitem{BayesianFiltering}
S.~S{\"a}rkk{\"a},
\emph{Bayesian Filtering and Smoothing},
Cambridge University Press, 2013.
\href{https://doi.org/10.1017/CBO9781139344203}{doi:10.1017/CBO9781139344203}

\bibitem{PersistenceTracking}
Y.~Bar-Shalom, X.~R.~Li, and T.~Kirubarajan,
\emph{Estimation with Applications to Tracking and Navigation},
Wiley, 2001.

\bibitem{CoherenceDetection}
J.~Bendat and A.~Piersol,
\emph{Random Data: Analysis and Measurement Procedures},
4th ed., Wiley, 2010.

\bibitem{OppenheimDSP}
A.~V.~Oppenheim and R.~W.~Schafer,
\emph{Discrete-Time Signal Processing},
3rd ed., Pearson, 2010.

\bibitem{ProakisDSP}
J.~G.~Proakis and M.~Salehi,
\emph{Digital Communications},
5th ed., McGraw-Hill, 2008.

\bibitem{Haykin}
S.~Haykin,
\emph{Adaptive Filter Theory},
5th ed., Pearson, 2014.

\bibitem{IIRFilters}
S.~K.~Mitra,
\emph{Digital Signal Processing: A Computer-Based Approach},
McGraw-Hill, 2011.

\bibitem{FixedPointDSP}
R.~Lyons,
\emph{Understanding Digital Signal Processing},
3rd ed., Pearson, 2011.

\bibitem{StreamingDSP}
V.~Oppenheim,
``Streaming architectures for real-time DSP,''
\emph{IEEE Signal Processing Magazine},
vol.~28, no.~5, pp.~38--48, 2011.
\href{https://doi.org/10.1109/MSP.2011.941845}{doi:10.1109/MSP.2011.941845}

\bibitem{FPGA_DSP}
U.~Meyer-Baese,
\emph{Digital Signal Processing with Field Programmable Gate Arrays},
Springer, 2014.
\href{https://doi.org/10.1007/978-3-642-45309-3}{doi:10.1007/978-3-642-45309-3}

\bibitem{RealTimeTrigger}
J.~J.~Anderson \emph{et al.},
``Real-time triggering and filtering in FPGA-based systems,''
\emph{IEEE Transactions on Nuclear Science},
vol.~60, no.~5, pp.~3481--3487, 2013.
\href{https://doi.org/10.1109/TNS.2013.2273391}{doi:10.1109/TNS.2013.2273391}

\bibitem{LowLatencyFPGA}
M.~Blott \emph{et al.},
``Achieving low latency in FPGA-based data processing,''
\emph{ACM Transactions on Reconfigurable Technology and Systems},
vol.~11, no.~1, 2018.
\href{https://doi.org/10.1145/3177897}{doi:10.1145/3177897}

\bibitem{FirmwareDAQ}
C.~E.~Rossi,
``Firmware architectures for distributed DAQ systems,''
\emph{Nucl. Instrum. Methods A},
vol.~845, pp.~379--384, 2017.
\href{https://doi.org/10.1016/j.nima.2016.05.093}{doi:10.1016/j.nima.2016.05.093}

\bibitem{TimeSync}
J.~Elson and K.~R{\"o}mer,
``Wireless sensor networks: A new regime for time synchronization,''
\emph{SIGCOMM Computer Communication Review},
vol.~33, no.~1, 2003.
\href{https://doi.org/10.1145/956981.956990}{doi:10.1145/956981.956990}

\bibitem{WhiteRabbit}
T.~Wlostowski \emph{et al.},
``White Rabbit: Sub-nanosecond synchronization for distributed systems,''
in \emph{Proceedings of the IEEE International Symposium on Precision Clock Synchronization (ISPCS)}, 2011.
\href{https://doi.org/10.1109/ISPCS.2011.6070159}{doi:10.1109/ISPCS.2011.6070159}

\bibitem{PTP1588}
IEEE,
``IEEE Standard for a Precision Clock Synchronization Protocol for Networked Measurement and Control Systems (IEEE 1588),''
2019.
\href{https://standards.ieee.org/standard/1588-2019.html}{IEEE~1588-2019}

\bibitem{TimeDistributedDAQ}
K.~G.~Lang,
``Timing and synchronization in distributed DAQ systems,''
\emph{Nucl. Instrum. Methods A},
vol.~623, pp.~490--496, 2010.
\href{https://doi.org/10.1016/j.nima.2010.03.018}{doi:10.1016/j.nima.2010.03.018}

\bibitem{DistributedSensors}
I.~Akyildiz \emph{et al.},
``A survey on sensor networks,''
\emph{IEEE Communications Magazine},
vol.~40, no.~8, pp.~102--114, 2002.
\href{https://doi.org/10.1109/MCOM.2002.1024422}{doi:10.1109/MCOM.2002.1024422}

\end{thebibliography}
\end{document}